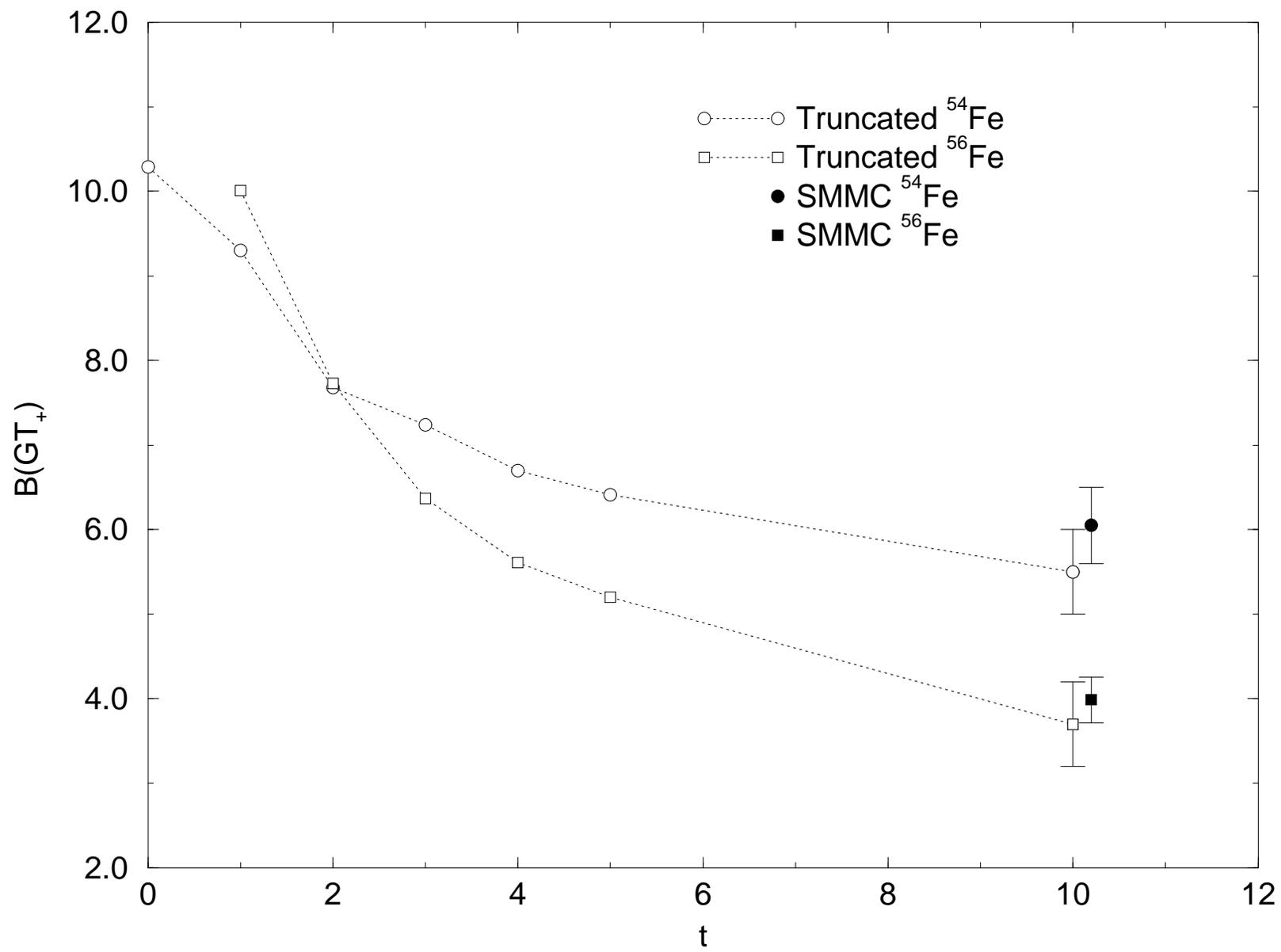

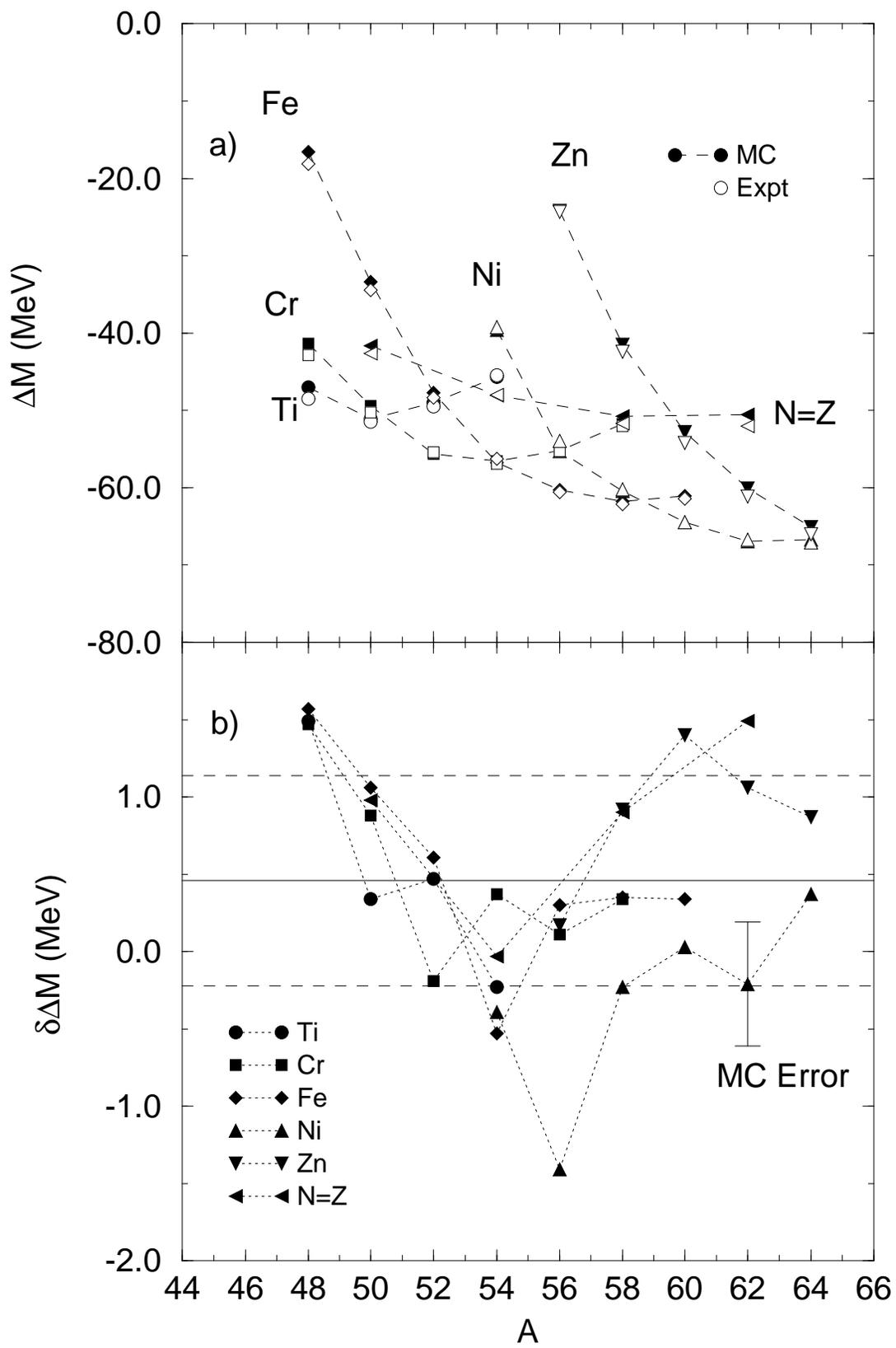

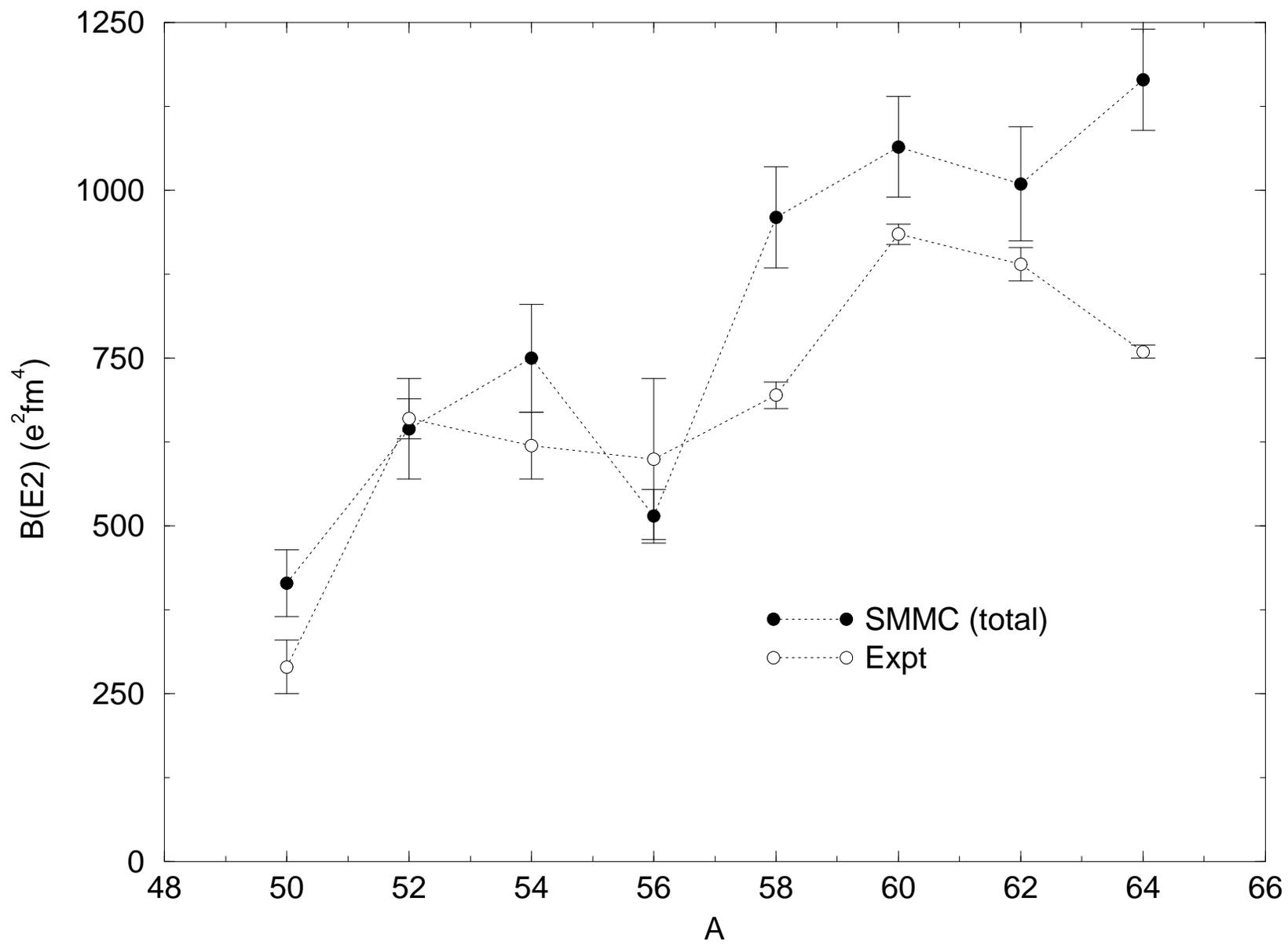

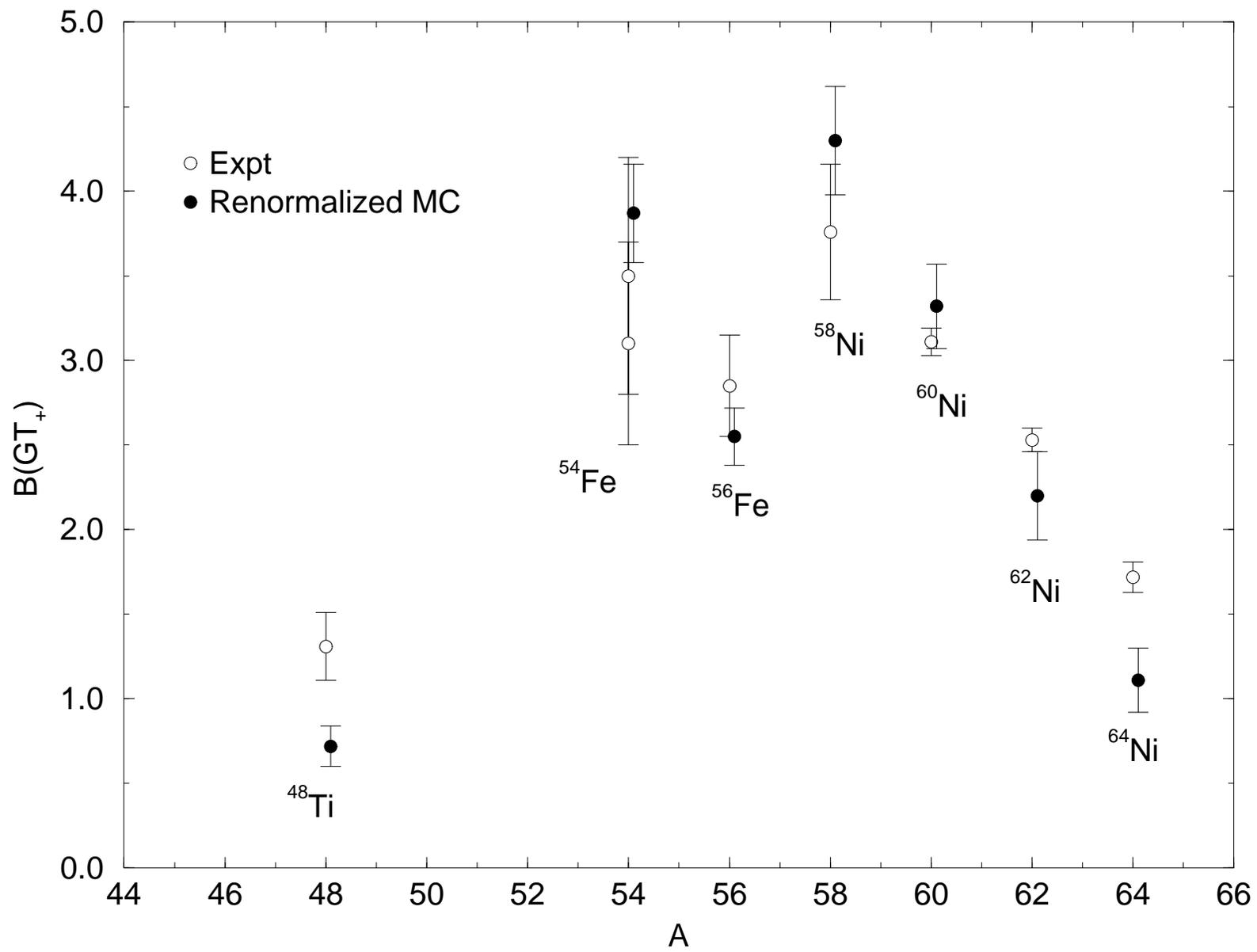

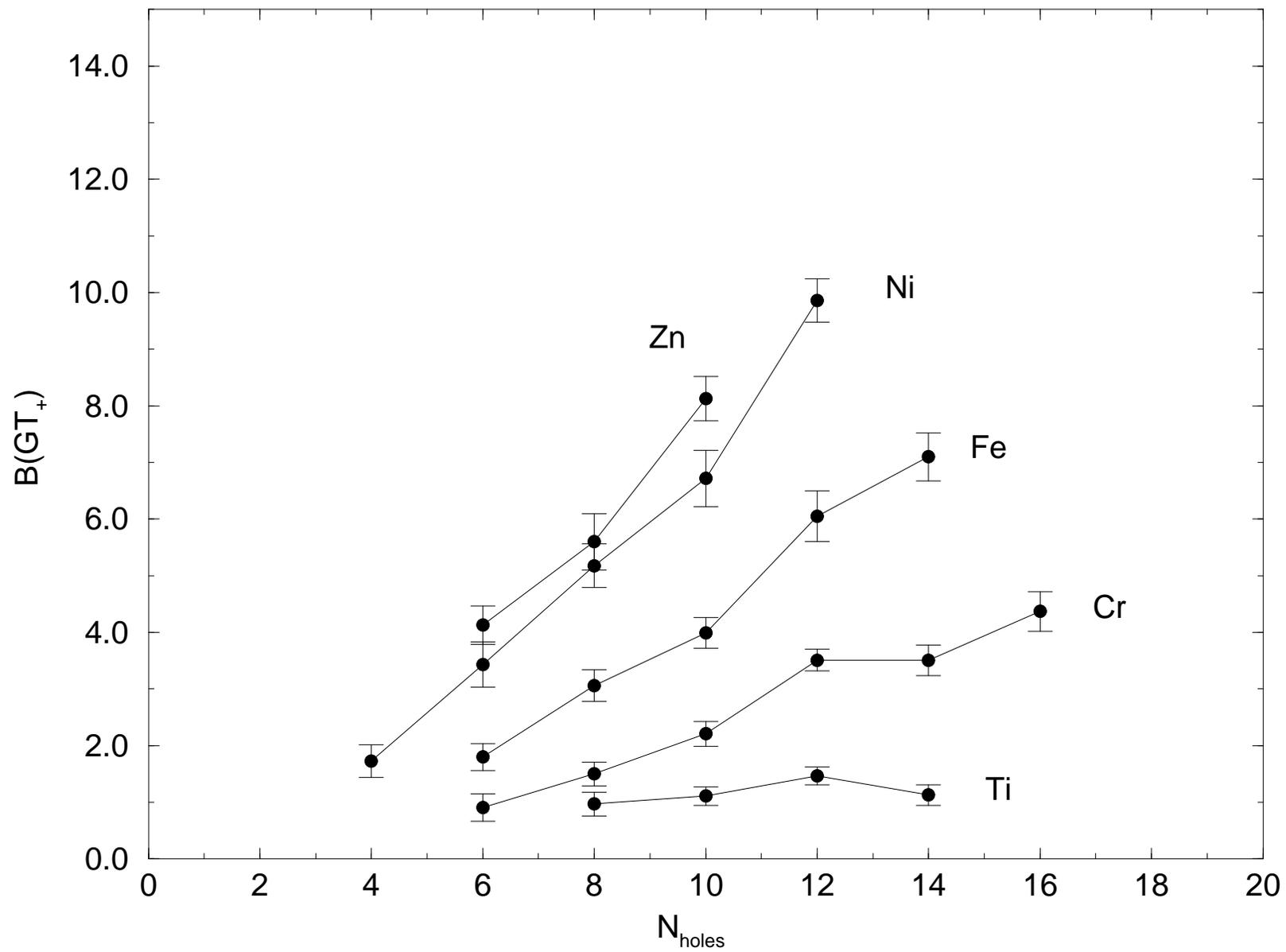

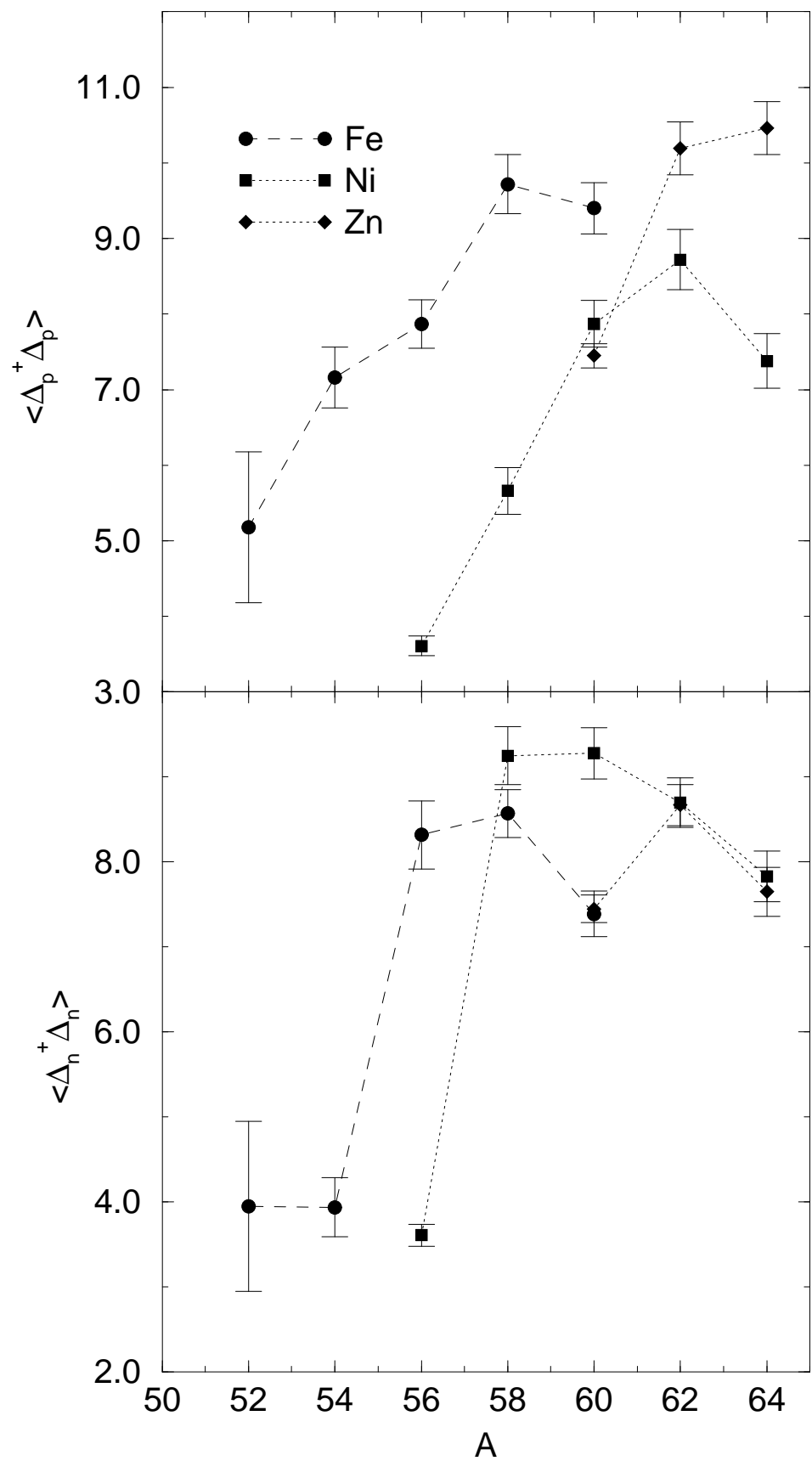

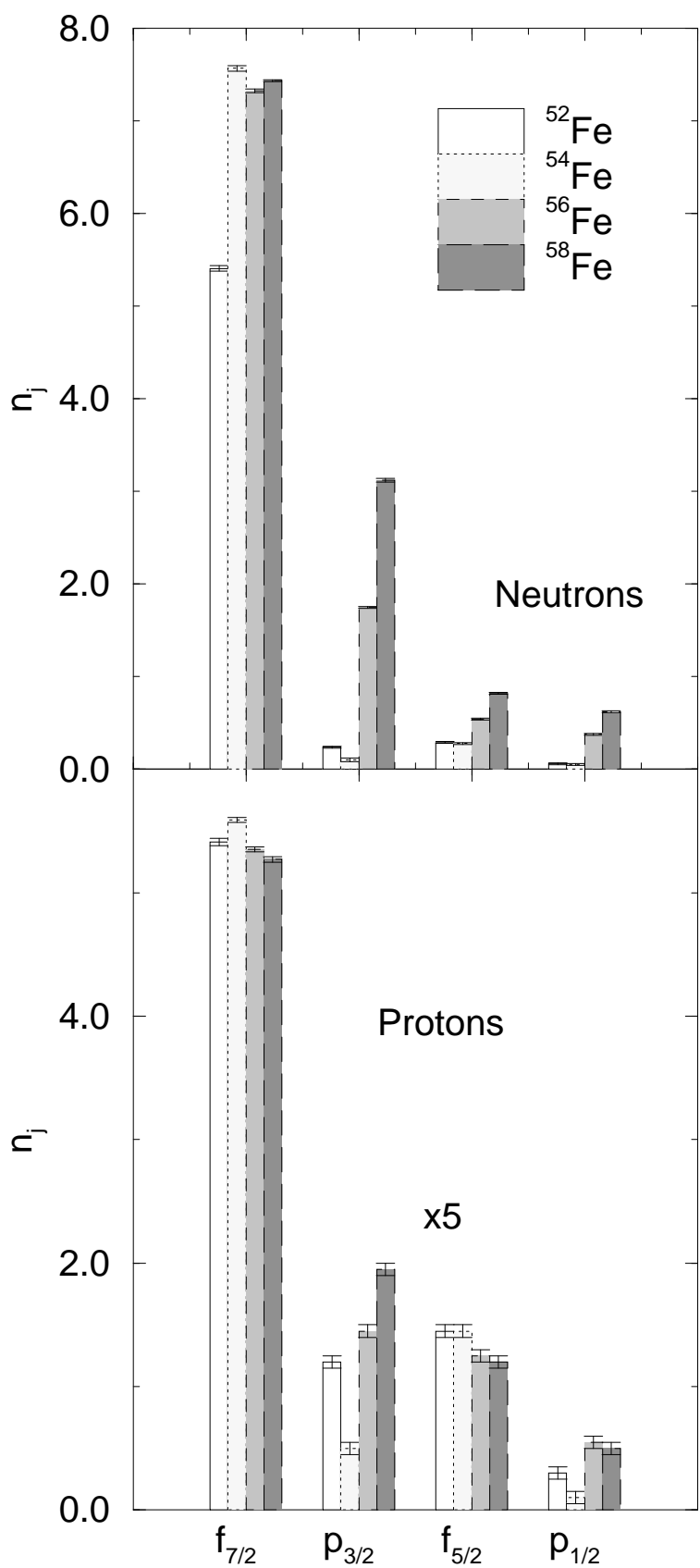

# Shell-model Monte Carlo studies of $fp$-shell nuclei

K. Langanke*, D. J. Dean*, P. B. Radha*, Y. Alhassid† and S. E. Koonin*

*W. K. Kellogg Radiation Laboratory, 106-38, California Institute of Technology, Pasadena, California 91125

† Center for Theoretical Physics, Sloane Physics Laboratory, Yale University, New Haven, Connecticut 06511

(April 19, 1995)

## Abstract

We study the gross properties of even-even and $N = Z$ nuclei with $A = 48 - 64$ using shell-model Monte Carlo methods. Our calculations account for all $0\hbar\omega$ configurations in the $fp$-shell and employ the modified Kuo-Brown interaction KB3. We find good agreement with data for masses and total $B(E2)$ strengths, the latter employing effective charges $e_p = 1.35e$ and $e_n = 0.35e$. The calculated total Gamow-Teller strengths agree consistently with the $B(GT_+)$-values deduced from $(n, p)$ data if the shell model results are renormalized by 0.64, as has already been established for $sd$-shell nuclei. The present calculations therefore suggest that this renormalization (i.e., $g_A = 1$ in the nuclear medium) is universal.

Typeset using REVTEX

# I. INTRODUCTION AND MOTIVATION

The interacting shell model [1] is generally considered to be the most fundamental theory of the nucleus short of an explicit solution of the $A$-body problem. Indeed, it is the conceptual basis for most other nuclear models. But in contrast to the atomic shell model, the residual interaction among the valence nucleons plays an important role. This interaction can be determined either by adjusting its matrix elements to fit a large volume of data, (as been done successfully for the $p$-shell [2] and $sd$ shell [3]) or from the nucleon-nucleon G-matrix [4].

Due to the combinatorial increase of configurations with the numbers of valence nucleons and orbitals, the conventional solution of the shell model by matrix diagonalization has been limited in heavier nuclei to truncated calculations. For example, complete $0\hbar\omega$ calculations in the $fp$-shell have been possible only for nuclei with $A \leq 48$ [5] and, very recently, for the ground state energy and Gamow-Teller strength of the nucleus $^{50}$Cr [6].

Conventional studies of $fp$-shell nuclei with $A \geq 50$ must be performed in severely truncated model spaces. This is unfortunate as, for example, the nuclei in the iron region play a crucial role in a supernova, when the Gamow-Teller strengths determine the electron capture rate and thus the dynamics of the early collapse. Truncated shell model calculations, however, recover only in part the well established Gamow-Teller quenching [7–9] and are thus inadequate for predicting the Gamow-Teller strengths of those nuclei, for which no experimental information is yet available.

Nevertheless, the conventional calculations performed for $A \leq 50$ support the conjecture that the shell model, if performed within a complete $0\hbar\omega$ basis, is able to describe the $fp$-shell nuclei. In particular, the systematic studies of the $A = 48$ nuclei [5] indicate that the residual KB3 interaction [10] is well suited to complete $fp$-shell calculations. This interaction has been derived by minimally modifying the monopole strength in the original Kuo-Brown interaction [4].

The results we present in this paper support that conjecture and demonstrate that com-plete $0\hbar\omega$ calculations performed with the KB3 interaction are capable of describing the ground-state properties of even-even and $N = Z$ nuclei in the $fp$-shell with $A \leq 64$ (for larger $A$ the $g_{9/2}$-orbital must be included). Our studies make use of the recently developed shell model Monte Carlo (SMMC) method [11] in which few-body observables are calculated at finite temperature. The SMMC exploits the fact that most of the billions of configurations in these nuclei are quite unimportant for general nuclear properties, so that only a subset of the relevant configurations need be sampled. In comparison with the conventional diagonalization method, the SMMC is not yet capable of providing detailed spectroscopic information. However, it has been shown that the SMMC approach is well-suited for studies of both ground state properties [11–13] (obtained in the low-temperature limit) and nuclei at finite temperature [14]. In this paper we systematically study important properties of $fp$-shell nuclei, including masses, and Gamow-Teller, $E2$, and $M1$ total strengths. As our results demonstrate, these calculations are in agreement with experiment for all these quantities over the wide range of nuclei studied ($A = 48 - 64$).

## II. THE SHELL MODEL MONTE CARLO METHOD

The shell model Monte Carlo method is based on a statistical formulation. The canonical expectation value of an observable $\mathcal{A}$ at a given temperature $T$ is given by [15,11,12,16] ($\beta = 1/T$)

$$\langle \mathcal{A} \rangle = \frac{\text{Tr}_A(\mathcal{A} e^{-\beta H})}{\text{Tr}_A(e^{-\beta H})}, \tag{1}$$

where $U = \exp(-\beta H)$ is the imaginary-time many-body propagator and $\text{Tr}_A U$ is the canonical partition function for $A$ nucleons. The shell model Hamiltonian $H$ can be cast in the form

$$H = \sum_\alpha \left( \epsilon_\alpha^* \bar{\mathcal{O}}_\alpha + \epsilon_\alpha \mathcal{O}_\alpha \right) + \frac{1}{2} \sum_\alpha V_\alpha \left\{ \mathcal{O}_\alpha, \bar{\mathcal{O}}_\alpha \right\}, \tag{2}$$

where $\epsilon_\alpha$ are the single particle energies and $\mathcal{O}_\alpha$ represent a set of one-body density operators ($\bar{\mathcal{O}}$ denotes the time-reverse of $\mathcal{O}$). The Hamiltonian in Eq. (2) is manifestly time-reversal invariant if the parameters $V_\alpha$ that define the strength of the residual two-body interactions are real.

The key to the SMMC method is to rewrite the propagator $U$ as a functional integral over one-body propagators. To achieve this goal, the exponent in $U$ is split into $N_t$ time slices of duration $\Delta\beta = \beta/N_t$,

$$U = \left[e^{-\Delta\beta H}\right]^{N_t}. \tag{3}$$

The many-body propagator at each time slice is linearized by a Hubbard-Stratonovich transformation [17]; *i.e.*, it is transformed into an integral over a set of one-body propagators that correspond to non-interacting nucleons in fluctuating auxiliary fields defined by complex c-numbers $\sigma_{\alpha n}$ ($n = 1, ..., N_t$). The expectation value of $\mathcal{A}$ then reads

$$\langle \mathcal{A} \rangle_A = \frac{\text{Tr}_A(\mathcal{A}e^{-\beta H})}{\text{Tr}_A(e^{-\beta H})} \approx \frac{\int D[\sigma] W(\sigma) \Phi(\sigma) \langle \mathcal{A} \rangle_\sigma}{\int D[\sigma] W(\sigma) \Phi(\sigma)}, \tag{4}$$

where the metric is

$$D[\sigma] = \Pi_{\alpha n}[d\sigma_{\alpha n} d\sigma^*_{\alpha n} \Delta\beta |V_\alpha|/2\pi], \tag{5}$$

and the approximation becomes exact as $N_t \to \infty$. The non-negative weight is

$$W(\sigma) = |\xi(\sigma)| \exp(-\frac{1}{2} \sum_{\alpha n} |V_\alpha| |\sigma_{\alpha n}|^2 \Delta\beta), \tag{6}$$

where $\xi(\sigma) = \text{Tr}_A U_\sigma$ is the partition function of the one-body propagator $U_\sigma = U_{N_t} \cdots U_1$, with $U_n = \exp(-\Delta\beta h_n)$, and the one-body Hamiltonian for the $n^{\text{th}}$ time slice is

$$h_n = \sum_\alpha (\epsilon^*_\alpha + s_\alpha V_\alpha \sigma_{\alpha n}) \bar{\mathcal{O}}_\alpha + (\epsilon_\alpha + s_\alpha V_\alpha \sigma^*_{\alpha n}) \mathcal{O}_\alpha, \tag{7}$$

with $s_\alpha = \pm 1$ for $V_\alpha < 0$ and $s_\alpha = \pm i$ for $V_\alpha > 0$. The "sign" is $\Phi(\sigma) = \xi(\sigma)/|\xi(\sigma)|$ and the expectation value of $\mathcal{A}$ with respect to the auxiliary field $\sigma$ is

$$\langle \mathcal{A} \rangle_\sigma = \text{Tr}_A \mathcal{A} U_\sigma / \xi(\sigma). \tag{8}$$

Both $\xi(\sigma)$ and $\langle \mathcal{A} \rangle_\sigma$ can be evaluated in terms of the matrix $\mathbf{U}_\sigma$ that represents the evolution operator $U_\sigma$ in the space of $N_s$ single-particle states. In the applications discussed below the trace is canonical corresponding to a nucleus with a fixed number of nucleons [16]. Details of the transformation from the residual particle-particle interaction to the $V_\alpha$ used above can be found in [11].

If all $V_\alpha < 0$, then the sign is $\langle \Phi \rangle = 1$. However, for realistic nuclear interactions such as KB3, about half of the $V_\alpha$'s are positive generating a sign problem (where the uncertainty in $\Phi$ is larger than $\langle \Phi \rangle$). To overcome this problem, we extrapolate observables calculated for a family of good-sign Hamiltonians $H_g$ (with $g < 0$) to the physical Hamiltonian at $g = 1$ [12].

The results presented in this paper correspond to various observables of the nuclear ground state: the energy $\langle H \rangle$, the total $B(M1)$ and $B(E2)$ strengths, and the Gamow-Teller strength. The total $B(M1)$ strength is defined as $B(M1) = \langle \vec{\mu}^2 \rangle$, where the magnetic moment $\vec{\mu}$ is given by $\vec{\mu} = \sum_i \mu_N \left\{ g_l \vec{l} + g_s \vec{s} \right\}$, $\mu_N$ is the nuclear magneton and $g_l$, $g_s$ are the free gyromagnetic ratios for angular momentum and spin, respectively ($g_l = 1$, $g_s = 5.586$ for protons, and $g_l = 0$, $g_s = -3.826$ for neutrons). The total $B(E2)$ strength is given by $B(E2) = \langle Q^2 \rangle$, where the quadrupole operator is defined by $Q = e_p Q_p + e_n Q_n$ with $Q_{p(n)} = e_{p(n)} \sum_i r_i^2 Y_2(\theta_i, \phi_i)$; the sum runs over all valence protons (neutrons). The effective charges $e_{p,n}$ account for coupling to the giant quadrupole resonance outside our model space. For the oscillator length, we used $b = 1.01 A^{1/6}$ fm. The total Gamow-Teller strength is given by $B(GT_\pm) = \langle (\vec{\sigma} \tau_\pm)^2 \rangle$. We also explore the isovector monopole pairing in the ground states, as described below.

The SMMC studies presented below have been performed in the complete set of $0f_{7/2,5/2} - 1p_{3/2,1/2}$ configurations using the modified Kuo-Brown KB3 residual interaction [10]. Each calculation involved 4000-5000 Monte Carlo samples at each of six values of the coupling constant $g$ equally spaced between $-1$ and $0$; extrapolation to the physical case ($g = 1$) was done by the method described in Ref. [14]. In many cases the assumption of a linear dependence on $g$ resulted in acceptable fits ($\chi^2$ per degree of freedom less than one). However, we have estimated the errors in our calculation conservatively by adopting a quadratic extrapolation in $g$. For the Hamiltonian we have made use of the variational prin-ciple which ensures that $\langle H \rangle$ has a minimum at $g = 1$ [12]. The calculations were performed at $\beta = 2$ MeV$^{-1}$ (which is sufficient to guarantee cooling within a few hundred keV of the ground state for even-even nuclei [14]), and $N_t = 64$ time slices (which results in negligible discretization errors).

## III. RESULTS

### A. Comparison with direct diagonalization

A few of the nuclei we consider here with SMMC methods have previously been studied by direct diagonalization using the same residual interaction. In all cases, the results obtained by these two very different methods agree.

Poves and collaborators calculated Gamow-Teller strengths of $B(GT_+) = 1.263$ for $^{48}$Ti, 4.13 for $^{48}$Cr [5], and 3.57 for $^{50}$Cr [6], while our SMMC calculations yield $1.13 \pm 0.18$, $4.37 \pm 0.35$ and $3.51 \pm 0.27$ for these nuclei, respectively. For the two iron isotopes $^{54}$Fe and $^{56}$Fe, Caurier et al. [6] calculated the Gamow-Teller strengths in a series of direct diagonalizations with a decreasing level of truncation. In Fig. 1 we compare the results of these truncation series with our full $0\hbar\omega$ SMMC results. Our results ($6.05 \pm 0.45$ for $^{54}$Fe and $3.99 \pm 0.27$ for $^{56}$Fe) agree with the values obtained by Caurier et al. upon extrapolation to no truncation ($5.5 \pm 0.5$ and $3.7 \pm 0.5$ for $^{54}$Fe and $^{56}$Fe, respectively). Fig. 1 also clearly demonstrates that complete ($0\hbar\omega$) shell model calculations recover significantly more quenching of the Gamow-Teller strength than truncated calculations. Note that the single particle estimate for the total Gamow-Teller strength in both nuclei is $B(GT_+) = 10.3$.

For the total $B(E2)$ strength from the ground state we can compare our results with those of conventional shell model calculations for the nuclei $^{48}$Ti [5] and $^{48}$Cr [18]. Using the effective charges $e_p = 1.5e$ and $e_n = 0.5e$ for protons and neutrons, respectively [5], these authors find $B(E2) = 583.5$ e$^2$fm$^4$ for $^{48}$Ti, while we calculate $630 \pm 55$ e$^2$fm$^4$. For later reference it is noteworthy that roughly 80% of the strength lies in the transition from the ground state to the first $2^+$ state [5], in agreement with experiment. For $^{48}$Cr, Poves et al. calculate the total $B(E2)$ strength of 1280 $e^2$fm$^4$ [18], in agreement with our SMMC result of $B(E2) = 1375 \pm 110$ $e^2$fm$^4$.

¿From Ref. [5] we deduce the energies $\langle H \rangle = -24.5$ MeV and $-32.9$ MeV for $^{48}$Ti and $^{48}$Cr, respectively. Our SMMC results for these quantities are $-23.9 \pm 0.4$ MeV and $-32.3 \pm 0.4$ MeV, respectively. For $^{50}$Cr we calculate $\langle H \rangle = -40.0 \pm 0.4$ MeV, while Caurier et al. [6] give $-40.57$ MeV. The slight underbinding we find is expected, as our calculations have been performed at the finite temperature $T = 0.5$ MeV and therefore should contain a small excitation energy of a few hundred keV [14].

More details of the comparison of SMMC calculations with exact diagonalization results with realistic interactions are given in Ref. [19].

### B. Energies

In this subsection we compare our calculated binding energies for the various nuclei with experimental data. The Coulomb energies, which are not included in the KB3 interaction, are calculated (in MeV) as [5] ($\pi$ = number of valence protons, $\nu$ = number of valence neutrons, $n = \pi + \nu$)

$$H_{\text{Coul}} = \frac{\pi(\pi-1)}{2} \cdot 0.35 - \pi\nu \cdot 0.05 + \pi \cdot 7.289. \qquad (9)$$

Following Ref. [5] we have added to the calculated energy expectation values an energy shift of $0.028 \cdot n(n-1)$ MeV to correct for a "tiny" residual monopole defect in the KB3 interaction. Note that the KB3 interaction uses the single particle energies from the experimental levels in $^{41}$Ca. Thus, our energy scale is set by $\epsilon = -8.363$ MeV for the $f_{7/2}$ orbital, obtained from the neutron separation energy of $^{41}$Ca.

The nuclear mass excesses $\Delta M$ obtained this way (relative to $^{40}$Ca) are compared to the data in Fig. 2a. The results obtained are very satisfactory, as can also be seen in Fig. 2b, which shows the discrepancy $\delta\Delta M$ between our calculation and data. Considering that our finite-temperature calculation includes a small internal excitation energy of a few hundred keV, the reproduction of the mass data by the shell model using the KB3 interaction is better than is suggested by Fig. 2b.

Two remarks can be made about our energy results. First, the accuracy of our choice of Coulomb energy (9) can be estimated by comparing the mass differences $\Delta M$ for the mirror nuclei ($^{48}$Fe, $^{48}$Cr), ($^{50}$Fe, $^{50}$Cr) and ($^{54}$Fe, $^{54}$Ni). We find deviations between 0.1 MeV ($A = 48$) and 0.5 MeV ($A = 54$) showing that our parametrization is sufficiently accurate, but might become less reliable with increasing $Z$. Second, in contrast to the other nuclei studied, the SMMC overbinds the Nickel isotopes, as well as $^{54}$Fe and $^{52}$Cr. As these nuclei have semi-magic proton or neutron numbers ($N = 28$), our results indicate that the KB3 interaction slightly overemphasizes the $N = 28$ shell closures. Indeed, the overbinding is strongest for the double-magic nucleus $^{56}$Ni.

### C. The $B(M1)$ and $B(E2)$ strengths

We have calculated the total $B(M1)$ strengths of the various nuclei using free nucleon g-factors, as listed in Table 1. Unfortunately the known $M1$ transitions in the $fp$-shell nuclei are mainly to low-lying states that exhaust only a small fraction of the total strength, so that a comparison to data is not possible in most cases. However, for a few of the nuclei studied here, Richter and collaborators [20–22] have used high-resolution electron scattering to study the $B(M1)$ strength by means of high-resolution electron scattering in an energy window large enough to contain most of the strength. For the $N = 28$ isotones (in units of $\mu_N^2$) they find $B(M1)= 4.5 \pm 0.5$ ($^{50}$Ti), $8.1 \pm 0.8$ ($^{52}$Cr) and $6.6 \pm 0.4$ ($^{54}$Fe) for excitation energies between 7 and 12 MeV [21]. If one considers that this energy window should contain about 75% of the total strength [21], our SMMC results ($B(M1) = 12.5 \pm 1.0$ ($^{50}$Ti), $18.9 \pm 2.2$ ($^{52}$Cr), $16.5 \pm 2.8$ ($^{54}$Fe)) are roughly twice the observed $B(M1)$ strength for these nuclei, supporting the idea that the spin g-factors are renormalized in the nuclear medium [23].

We note that both the independent particle and $1p - 1h$ shell models, discussed in Ref.

[21] predict a linear rise of the total $B(M1)$ strength with the number of valence protons in the $N = 28$ isotones. Our SMMC calculation does not support this trend, as the calculated $B(M1)$ strength in $^{54}$Fe is no larger than that for $^{52}$Cr, in agreement with experiment [21]. We also note that the SMMC results might be tested by comparing the isotone pairs ($^{52}$Cr,$^{54}$Fe) and ($^{54}$Cr,$^{56}$Fe) – the calculations predict roughly the same $B(M1)$ strength in the Iron isotopes, but a significantly smaller $B(M1)$ strength in $^{54}$Cr than in $^{52}$Cr.

For $^{58}$Ni Mettner *et al.* determined a total $B(M1)$ strength of $16.9^{+4.6}_{-3.3}$, while the SMMC result is $20 \pm 2$. Within the large (experimental and theoretical) errors, these results are compatible with a significant renormalization of the (spin) g-factors.

Table 1 also lists our calculated total $B(E2)$ strength for transitions from the ground state. We have calculated these quantities by adopting the same effective charges used in a recent truncated shell model calculation of $^{54}$Fe [7] ($e_p = 1.35e$ and $e_n = 0.35e$). For comparison Table 1 also lists the measured $B(E2)$ values for the $0_1^+ \to 2_1^+$ transitions in these nuclei; in even-even nuclei this transition typically exhausts about $70 - 80\%$ of the total strength. For example, from the $(e,e')$ data [28] we calculate ratios between the total observed $B(E2)$ strength and the experimental $B(E2, 0_1^+ \to 2_1^+)$ by factors of 1.35, 1.14 and 1.20 for the Nickel isotopes $^{58,60,64}$Ni, respectively, while our SMMC results exceed the experimental $B(E2, 0_1^+ \to 2_1^+)$ values by 1.38, 1.14 and 1.53. The overall level of agreement is illustrated in Fig. 3, where we compare the total calculated $B(E2)$ strength with the experimenal $B(E2, 0_1^+ \to 2_1^+)$ values for those $fp$-shell nuclei with either semi-magic proton or neutron number $N = 28$. A similar comparison has been presented in Ref. [29] based on a strongly truncated shell model calculation and a different residual interaction. To achieve overall agreement with the data, Ref. [29] used somewhat larger effective charges ($e_p = 1.4e$ and $e_n = 0.9e$) in order to compensate for correlations missing in the truncated model space. Note that the use of the larger effective charges would increase our total $B(E2)$ strengths for the nickel isotopes by more than 60%.

While our calculation apparently compares nicely with data for the nuclei in the middle of the shell, our $B(E2)$ values as listed in Table 1 appear too low for some Chromium ($^{48,50}$Cr) and Titanium isotopes ($^{48}$Ti). For these nuclei larger effective charges are required, perhaps indicating a greater importance of $sd$-shell configurations at the beginning of the shell.

### D. Gamow-Teller strength

In previous publications [12,13] we have shown that full $0\hbar\omega$ shell model calculations recover significantly more quenching in $fp$-shell nuclei than truncated $2p-2h$ calculations; a finding that is in agreement with the recent work of the Strasburg-Madrid group [5,6]. Our work suggested that isoscalar proton-neutron correlations [30] and proton and neutron pairs coupled to non-zero angular momenta [31] are mainly responsible for the quenching of the Gamow-Teller strength in the ground states. However, we also observed, by performing calculations for two different residual interactions (the Brown-Richter interaction and the original Kuo-Brown interaction), that the calculated Gamow-Teller strength is rather sensitive to the residual interaction [13]. Moreover no systematic trends between the calculations using these forces and the data could be detected. As we demonstrate in the following, the situation improves significantly if one employs the modified Kuo-Brown interaction KB3.

Our results for the total Gamow-Teller strengths $B(GT_+) = \langle(\vec{\sigma}\tau_+)^2\rangle$ are listed in Table 1. (As our calculation obeys the Ikeda sum rule, values for $B(GT_-)$ are readily obtained by adding $3(N-Z)$ to the $B(GT_+)$ values.) We observe that the calculated values are systematically larger than the $B(GT_+)$ values deduced from intermediate-energy $(n,p)$ charge-exchange cross section data at forward angles, which are known to be dominated by the $GT_+$ operator and currently provide the only experimental determination of the Gamow-Teller strength function. However, such a systematic overestimation of the Gamow-Teller strength by shell model calculations is familiar from work in the $sd$-shell and can be attributed to an in-medium renormalization of the axial-vector coupling constant $g_A$ [3]. Since the $(n,p)$ data are usually normalized to low-energy beta-decay rates, they are therefore indirectly also subject to any renormalization of $g_A$. To account for this renormalization the shell model results are usually multiplied by $(g_A^{\text{eff}}/g_A)^2 = (1./1.25)^2 = 0.64$ [5,6]. If we apply this rescaling to our $B(GT_+)$ results for the $fp$-shell we find good agreement between the SMMC calculations and the data, as is shown in Fig. 4. For $^{48}$Ti and $^{64}$Ni our renormalized $B(GT_+)$ values deviate slightly from the measured Gamow-Teller strengths, indicating possible limits of the present model space. The slight discrepancy in the case of $^{48}$Ti might be of some importance for shell model calculations of the double-beta decay rate of $^{48}$Ca [32].

¿From Fig. 4 we conclude that i) Full $0\hbar\omega$ shell model calculations describe the systematics of the Gamow-Teller quenching in the $fp$-shell; ii) Reproduction of the data requires a renormalization by 0.64, in agreement with the usual assumption of an in-medium modification of $g_A$; and iii) these results for the $fp$-shell are consistent with those deduced previously for the $sd$-shell and indicate that both conclusions i) and ii) might be universal. If one accepts these conclusions, the agreement between data and theory suggests both that the $(n,p)$ experiments do not miss any significant strength at higher energies and that the KB3 interaction well-describes isoscalar correlations in $fp$-shell nuclei.

The unrenormalized $B(GT_+)$ strengths for the various isotopic chains are plotted in Fig. 5. As predicted by the simple single-particle estimate [33,34], the $B(GT_+)$ strength is roughly constant in the titanium isotopes, corresponding to a quenching factor of 3. For the other isotopic chains our calculation is in agreement with a recently suggested systematics of the experimental data [34] in which the total $B(GT_+)$ strength for mid-$fp$-shell nuclei is proportional to the numbers of valence protons and neutron holes in the $fp$-shell. As can be seen in Fig. 5, the linear proportionality to the number of neutron holes is also found in our Monte Carlo results. We also observe that the Gamow-Teller strength in the nuclei with 12 neutron holes, corresponding to the magic neutron number $N = 28$, is larger than the trend in the nuclei of the same isotope chain, which might further indicate an overestimate of the shell closure by the KB3 interaction. ¿From the slopes of the $B(GT_+)$ strengths for the various isotope chains we find that $B(GT_+)$ scales with the number of valence protons, as expected.

For the even-even $N = Z$ nucleus $^{64}$Ge we find an unrenormalized total Gamow-Teller strength of $B(GT_+) = 7.91 \pm 0.54$. For the odd-odd $N = Z$ nuclei we calculate $B(GT_+)$ values of $8.1 \pm 2.5$ ($^{50}$Mn), $9.1 \pm 1.7$ ($^{54}$Co), $6.6 \pm 2.8$ ($^{58}$Cu) and $9.1 \pm 2.2$ ($^{62}$Ga). The large uncertainties prohibit us from drawing any meaningful conclusions from these values.

### E. Proton-proton and neutron-neutron pairing

It is well known that pairing between like nucleons plays an essential role for the ground state properties of even-even nuclei. In a first approximation the pairing can be described by the BCS model, which assumes that like nucleons are coupled to $J = 0$ pairs. We have studied the BCS-like pairing content of the ground states by measuring the expectation values for the pairing fields, $\langle \Delta^\dagger \Delta \rangle$, for proton-proton and neutron-neutron pairing. Here the pair operator is defined as $\Delta^\dagger = \sum_{j,m>0} a^\dagger_{jm} a^\dagger_{j\bar{m}}$, where $\bar{m}$ is the time-reverse orbit of $m$. For a Fermi gas with occupations $n_j = \sum_m \langle a^\dagger_{jm} a_{jm} \rangle$ we note that

$$\langle \Delta^\dagger \Delta \rangle = \sum_j \frac{n_j^2}{2(2j+1)}. \tag{10}$$

The results, obtained after subtraction of the independent particle model values (10), are shown in Fig. 6 for three different isotope chains. As expected, the excess pair correlations are generally quite strong and exceed the independent particle model values by factors of 2 to 4, reflecting the known coherence in the ground states of even-even nuclei. It is interesting to note that the proton pairing fields are not constant within an isotope chain, but usually increase with neutron number, showing that there are important neutron-proton correlations present in these ground states. The shell closure at $N = 28$ is manifest in the neutron pairing. For $^{52,54}$Fe and $^{56}$Ni the neutrons prefer to take advantage of closing the $f_{7/2}$-subshell, making the excess neutron pairing rather small (about a factor of two compared to the independent particle model). But once there are extra neutrons outside the closed subshell, the excess of pairing increases drastically. As is demonstrated in Fig. 7 these strong changes are not present in the average occupation numbers $n_j$. For example, the proton occupation numbers show little variation within the iron chain $^{52-58}$Fe, although the occupation of the $f_{7/2}$ orbit is largest in the semi-magic nucleus $^{54}$Fe. A similarly smooth behaviour is found in the neutron occupation numbers, where the additional neutrons are, on average, added to the lowest possible orbital. It is interesting to note that, in $^{52,54}$Fe, as a result of a strong $f_{7/2}$-$f_{5/2}$ coupling, the occupation of the energetically unfavored $f_{5/2}$ orbital is larger than that of the $p_{3/2}$ orbitals. The presence of neutron pairs in the $p_{3/2}$ orbital, as in $^{56,58}$Fe, also increases the occupation number for protons in this orbital, at the expense of the occupation of the ground state orbital. The comparison of Figs. 6 and 7 is yet another example for the importance of nucleon correlations in the ground state beyond the mean field level.

## IV. CONCLUSIONS

Using the recently developed shell model Monte Carlo approach we have studied the gross properties of even-even nuclei in the mid-$fp$ shell within full $0\hbar\omega$ shell model calculations. Our studies use the KB3 interaction, a minimally corrected version of the original Kuo-Brown G-matrix interaction. Conventional diagonalization approaches have already established that shell model calculations using this KB3 interaction give a very satisfying description of nuclei at the beginning of the $fp$-shell [5,6]. The present calculations supplement these studies and show that shell model calculations, using this physical interaction, satisfyingly describe the gross properties of (even-even) nuclei throughout that part of the $fp$-shell ($A \leq 64$) where the influence of the $g_{9/2}$ orbital should still be small. We view this reproduction of such a large body of data, as presented in this paper, as a remarkable success of a microscopic model.

We recall that performing full ($0\hbar\omega$) calculations of mid-$fp$-shell nuclei, as discussed in this paper, was unthinkable only two years ago. Thus, the present studies and their results are also a successful confirmation of the Monte Carlo approach to the nuclear shell model, and establish it as a powerful tool with which to study gross nuclear properties. It is certainly a useful complement to the conventional diagonalization approaches, which remain the method of choice for detailed spectroscopic properties.

Our calculations have shown that shell model calculations with the KB3 interaction describe the binding energies of even-even and $N = Z$ nuclei with $A = 48 - 64$ to within 1 MeV or better. We are quite aware that shell model calculations often do even better. It is conceivable that an improved reproduction of experimental masses requires the correction of tiny residual monopole defects in the KB3 interaction, as suggested in Ref. [5]. Such an optimization of the interaction is beyond the scope of the present calculation, but is conceivable with present computer capabilities. Our studies also indicate that the KB3 interaction slightly overemphasizes the $N = 28$ shell closure.

Using conventional effective charges that account for coupling to the giant quadrupole resonance outside our model space, our calculated $B(E2)$ strengths apparently reproduce the trend in the data suggested by the experimental $B(E2)$ values for the transition from the ground state to the first $2^+$ state. This transition exhaust typically $70-80\%$ of the total $B(E2)$ strength in even-even mid-$fp$-shell nuclei.

The most important result of our present study is a clarification of the Gamow-Teller quenching puzzle in the astrophysically important mid-$fp$-shell region, which has been the focus of attention for several years. While several mechanisms for the Gamow-Teller quenching have already been identified [30,31,13], we are, for the first time, able to demonstrate that the experimentally observed quenching is consistently reproduced by the correlations within the full $fp$-shell, if the standard renormalization factor of 0.64 is invoked. We find that the complete $0\hbar\omega$ calculations recover significantly more quenching than truncated shell model studies. Our results are consistent with previous ($0\hbar\omega$) calculations of $sd$-shell nuclei [3] which established that the Gamow-Teller strength is quenched by an additional renormalization factor 0.64 beyond the many-body correlations within that full model space. Thus, this additional renormalization of the Gamow-Teller strength appears to be universal and likely originates outside of nuclear configuration mixing within one major shell; it is consistent with $g_A = 1$ in the nuclear medium.

This work was supported by NSF grants PHY91-15574, PHY94-12818 and PHY94-20470 and by DOE grant DE-FG02-91-ER40608. We thank Prof. A. Poves for useful discussions and for supplying us with some results of his shell model diagonalizations prior to publication. Computational cycles were provided by the Concurrent Supercomputing Consortium and by the VPP500, a vector parallel processor at the RIKEN supercomputing facility; we thank Drs. I. Tanihata and S. Ohta for their assistance with the latter.

TABLES

TABLE I. Total $B(M1)$ (in $\mu_N^2$), $B(E2)$ (in $e^2 fm^4$) and unrenormalized $B(GT_+)$ strengths as calculated in the SMMC approach. For comparison the experimental $B(GT_+)$ strengths [24–26,9] and the $B(E2)$ values for the $0_1^+ \to 2_1^+$ transition [27,29] are also given.

| nucleus | $B(M1)$ | $B(E2)$(SMMC) | $B(E2)$ (exp) | $B(GT_+)$ (SMMC) | $B(GT_+)$ (exp) |
|---|---|---|---|---|---|
| $^{48}$Ti | 10.2 ± 1.2 | 455 ± 25 | 720 ± 40 | 1.13 ± 0.18 | 1.31±0.2 |
| $^{50}$Ti | 12.5 ± 1.0 | 415 ± 50 | 290 ± 40 | 1.47 ± 0.16 | |
| $^{52}$Ti | 12.5 ± 1.0 | 465 ± 55 | ≥ 250 | 1.11 ± 0.16 | |
| $^{54}$Ti | 13.5 ± 1.5 | 450 ± 80 | | 0.97 ± 0.21 | |
| $^{48}$Cr | 13.8 ± 1.7 | 945 ± 45 | 1330±200 | 4.37 ± 0.35 | |
| $^{50}$Cr | 14.5 ± 2.5 | 890 ±90 | 1080±60 | 3.51 ± 0.27 | |
| $^{52}$Cr | 18.9 ± 2.2 | 645 ±75 | 660±30 | 3.51 ± 0.19 | |
| $^{54}$Cr | 13.0 ± 2.5 | 890 ±90 | 870±40 | 2.21 ± 0.22 | |
| $^{56}$Cr | 16.2 ± 2.0 | 840 ±90 | | 1.50 ± 0.21 | |
| $^{52}$Fe | 18.9 ± 1.4 | 1055 ± 50 | | 7.10 ± 0.42 | |
| $^{54}$Fe | 16.5 ± 2.8 | 750 ± 80 | 620±50 | 6.05 ± 0.45 | 3.1±0.6 |
| $^{56}$Fe | 20.4 ± 3.0 | 990 ± 65 | 980±40 | 3.99 ± 0.27 | 2.85±0.3 |
| $^{58}$Fe | 20.3 ± 3.0 | 1010 ± 65 | 1200±40 | 3.06 ± 0.28 | |
| $^{60}$Fe | 17.3 ± 2.1 | 1105 ± 65 | 930±180 | 1.80 ± 0.24 | |
| $^{56}$Ni | 22.5 ± 1.2 | 515 ± 40 | 600±120 | 9.86 ± 0.38 | |
| $^{58}$Ni | 20.0 ± 2.0 | 960 ± 75 | 695±20 | 6.72 ± 0.50 | 3.76±0.4 |
| $^{60}$Ni | 22.0 ± 2.5 | 1065 ± 75 | 935±15 | 5.18 ± 0.39 | 3.11±0.08 |
| $^{62}$Ni | 19.6 ± 2.9 | 1010 ± 85 | 890±25 | 3.43 ± 0.40 | 2.53±0.07 |
| $^{64}$Ni | 18.9 ± 2.7 | 1165 ± 75 | 760±80 | 1.73 ± 0.29 | 1.72±0.09 |
| $^{60}$Zn | 19.5 ± 1.2 | 1335 ± 50 | | 8.13 ± 0.39 | |
| $^{62}$Zn | 19.0 ± 2.2 | 1350 ± 70 | 1230±90 | 5.60 ± 0.50 | |
| $^{64}$Zn | 23.6 ± 2.2 | 1225 ± 65 | 1440±120 | 4.13 ± 0.34 | |

FIGURES

FIG. 1. Comparison of the total Gamow-Teller strengths $B(GT_+)$ for $^{54,56}$Fe in a series of direct diagonalizations with decreasing level of truncation [6] with present full $fp$-shell result (solid symbols plotted at $t = 10$). The open symbols at $t = 10$ show the extrapolated no-truncation result of [6].

FIG. 2. Upper panel (a): Comparison of the mass excesses $\Delta M$ as calculated within the SMMC approach with the data. Lower panel (b): Discrepancy between the SMMC results for the mass excesses and the data, $\delta\Delta M$. The solid line shows the average discrepancy, 450 keV, while the dashed lines show the rms variation about this value.

FIG. 3. Comparison of experimental $B(E2, 0_1^+ \to 2_1^+)$ strengths with the total $B(E2)$ strength calculated in the SMMC approach for various $fp$-shell nuclei with either proton or neutron number $N = 28$.

FIG. 4. Comparison of the renormalized total Gamow-Teller strength, as calculated within the present SMMC approach, and the experimental $B(GT_+)$ values, deduced from $(n,p)$ data [24–26,9,35]. Note that the two measurements of $B(GT_+)$ for $^{54}$Fe summed the strength up to 8 MeV ($3.1 \pm 0.6$, [25]) and up to 9 MeV ($3.5 \pm 0.7$, [35]).

FIG. 5. Unrenormalized total Gamow-Teller strength for various isotope chains as a function of the number of neutron holes $(N - 20)$ in the $fp$-shell.

FIG. 6. Expectation values for the proton (upper) and neutron (lower) pairing fields, as calculated in the even iron, nickel and zinc isotopes. The values of $\langle\Delta^\dagger\Delta\rangle$ for the independent particle model have been subtracted.

FIG. 7. Occupation numbers of the various orbitals in the iron isotopes $^{52-58}$Fe, as calculated in the present SMMC approach. For clarity the proton occupation numbers of the $p_{3/2}$, $f_{5/2}$, and $p_{1/2}$ orbitals have been multiplied by a factor of 5.